\PassOptionsToPackage{unicode}{hyperref}
\PassOptionsToPackage{hyphens}{url}
\PassOptionsToPackage{dvipsnames,svgnames,x11names}{xcolor}
\documentclass[
]{article}
\usepackage{xparse}
\usepackage{amsmath,amssymb}
\usepackage{lmodern}
\usepackage{iftex}
\ifPDFTeX
  \usepackage[T1]{fontenc}
  \usepackage[utf8]{inputenc}
  \usepackage{textcomp} % provide euro and other symbols
\else % if luatex or xetex
  \usepackage{unicode-math}
  \defaultfontfeatures{Scale=MatchLowercase}
  \defaultfontfeatures[\rmfamily]{Ligatures=TeX,Scale=1}
\fi
\IfFileExists{upquote.sty}{\usepackage{upquote}}{}
\IfFileExists{microtype.sty}{% use microtype if available
  \usepackage[]{microtype}
  \UseMicrotypeSet[protrusion]{basicmath} % disable protrusion for tt fonts
}{}
\makeatletter
\@ifundefined{KOMAClassName}{% if non-KOMA class
  \IfFileExists{parskip.sty}{%
    \usepackage{parskip}
  }{% else
    \setlength{\parindent}{0pt}
    \setlength{\parskip}{6pt plus 2pt minus 1pt}}
}{% if KOMA class
  \KOMAoptions{parskip=half}}
\makeatother
\usepackage{xcolor}
\usepackage{color}
\usepackage{fancyvrb}

\DefineVerbatimEnvironment{Highlighting}{Verbatim}{commandchars=\\\{\}}
\newenvironment{Shaded}{}{}

\newcommand{\BuiltInTok}[1]{\textcolor[rgb]{0.00,0.50,0.00}{#1}}

\newcommand{\CommentTok}[1]{\textcolor[rgb]{0.38,0.63,0.69}{\textit{#1}}}

\newcommand{\ControlFlowTok}[1]{\textcolor[rgb]{0.00,0.44,0.13}{\textbf{#1}}}

\newcommand{\DecValTok}[1]{\textcolor[rgb]{0.25,0.63,0.44}{#1}}

\newcommand{\FloatTok}[1]{\textcolor[rgb]{0.25,0.63,0.44}{#1}}

\newcommand{\ImportTok}[1]{\textcolor[rgb]{0.00,0.50,0.00}{\textbf{#1}}}

\newcommand{\KeywordTok}[1]{\textcolor[rgb]{0.00,0.44,0.13}{\textbf{#1}}}
\newcommand{\NormalTok}[1]{#1}
\newcommand{\OperatorTok}[1]{\textcolor[rgb]{0.40,0.40,0.40}{#1}}

\newcommand{\StringTok}[1]{\textcolor[rgb]{0.25,0.44,0.63}{#1}}
\newcommand{\VariableTok}[1]{\textcolor[rgb]{0.10,0.09,0.49}{#1}}

\usepackage{graphicx}
\makeatletter
\def\maxwidth{\ifdim\Gin@nat@width>\linewidth\linewidth\else\Gin@nat@width\fi}
\def\maxheight{\ifdim\Gin@nat@height>\textheight\textheight\else\Gin@nat@height\fi}
\makeatother
\setkeys{Gin}{width=\maxwidth,height=\maxheight,keepaspectratio}
\makeatletter
\def\fps@figure{htbp}
\makeatother
\NewDocumentCommand\citeproctext{}{}
\NewDocumentCommand\citeproc{mm}{%
  \begingroup\def\citeproctext{#2}\cite{#1}\endgroup}
\makeatletter
 \let\@cite@ofmt\@firstofone
 \def\@biblabel#1{}
 \def\@cite#1#2{{#1\if@tempswa , #2\fi}}
\makeatother
\newlength{\cslhangindent}
\newlength{\csllabelwidth}
\newenvironment{CSLReferences}[2] % #1 hanging-indent, #2 entry-spacing
 {\begin{list}{}{%
  \setlength{\itemindent}{0pt}
  \setlength{\leftmargin}{0pt}
  \setlength{\parsep}{0pt}
  \ifodd #1
   \setlength{\leftmargin}{\cslhangindent}
   \setlength{\itemindent}{-1\cslhangindent}
  \fi
  \setlength{\itemsep}{#2\baselineskip}}}
 {\end{list}}
\usepackage{calc}

\ifLuaTeX
\usepackage[bidi=basic]{babel}
\else
\usepackage[bidi=default]{babel}
\fi
\babelprovide[main,import]{american}
\def\languageshorthands#1{}
\ifLuaTeX
  \usepackage{selnolig}  % disable illegal ligatures
\fi
\IfFileExists{bookmark.sty}{\usepackage{bookmark}}{\usepackage{hyperref}}
\IfFileExists{xurl.sty}{\usepackage{xurl}}{} % add URL line breaks if available
\hypersetup{
  pdftitle={PySLSQP: A transparent Python package for the SLSQP
optimization algorithm modernized with utilities for visualization and
post-processing},
  pdfauthor={Anugrah Jo Joshy, John T. Hwang},
  pdflang={en-US},
  colorlinks=true,
  linkcolor={Maroon},
  filecolor={Maroon},
  citecolor={Blue},
  urlcolor={Blue},
  pdfcreator={LaTeX via pandoc}}

\title{\texttt{PySLSQP}: A transparent Python package for the SLSQP
optimization algorithm modernized with utilities for visualization and
post-processing}

\usepackage[affil-it]{authblk}
\usepackage{orcidlink}
\author[1%
  ]{Anugrah Jo Joshy%
    \,\orcidlink{0009-0003-7704-2532}\,%
    }
\author[2%
  ]{John T. Hwang%
    }

\affil[1]{PhD Candidate, Department of Mechanical and Aerospace
Engineering, University of California San Diego}
\affil[2]{Associate Professor, Department of Mechanical and Aerospace
Engineering, University of California San Diego}
\date{6 August 2024}

\begin{document}
\maketitle

\section{Summary}\label{summary}

Nonlinear programming (NLP) addresses optimization problems involving
nonlinear objective and/or constraint functions defined over continuous
optimization variables. These functions are assumed to be smooth, i.e.,
continuously differentiable. Nonlinear programming has applications
ranging from aircraft design in engineering to optimizing portfolios in
finance and training models in machine learning. Sequential Quadratic
Programming (SQP) is one of the most successful classes of algorithms
for solving NLP problems. It solves an NLP problem by iteratively
formulating and solving a sequence of Quadratic Programming (QP)
subproblems. The Sequential Least SQuares Programming algorithm, or
SLSQP, has been one of the most widely used SQP algorithms since the
1980s.

We present \texttt{PySLSQP}, a seamless interface for using the SLSQP
algorithm from Python, that wraps the original Fortran code sourced from
the SciPy repository and provides a host of new features to improve the
research utility of the original algorithm. \texttt{PySLSQP} uses a
simple yet modern workflow for compiling and using Fortran code from
Python. This allows even beginner developers to easily modify the
algorithm in Fortran for their purposes and use in Python the wrapper
auto-generated by the workflow.

Some of the additional features offered by \texttt{PySLSQP} include
auto-generation of unavailable derivatives using finite differences,
independent scaling of the problem variables and functions, access to
internal optimization data, live-visualization, saving optimization data
from each iteration, warm/hot restarting of optimization, and various
other utilities for post-processing.

\texttt{PySLSQP} solves the general nonlinear programming problem:

\[
\begin{array}{rlr}
\underset{x \in \mathbb{R}^n}{\text{minimize}} & \; \; f(x) & \\
\text{subject to} & \begin{array}{ll}
                      c_i(x) = 0, &\quad i = 1,...,m_{eq} \\
                      c_i(x) \geq 0, &\quad i = m_{eq}+1,...,m \\
                      l_i \leq x_i \leq u_i, &\quad i = 1,...,n
                    \end{array}
\end{array}
\]

where \(x \in \mathbb{R}^n\) is the vector of optimization variables,
\(f: \mathbb{R}^n \to \mathbb{R}\) is the objective function,
\(c: \mathbb{R}^n \to \mathbb{R}^m\) is the vector-valued constraint
function, and \(l\) and \(u\) are the vectors containing the lower and
upper bounds for the optimization variables, respectively. The first
\(m_{eq}\) constraints are equalities while the remaining
\((m - m_{eq})\) constraints are inequalities.

\section{Statement of need}\label{statement-of-need}

The original SLSQP algorithm (\citeproc{ref-kraft1988software}{Kraft,
1988}, \citeproc{ref-kraft1994algorithm}{1994}), implemented in Fortran
by Dieter Kraft, has been incorporated into several software packages
for optimization across different programming languages. However, the
algorithm itself has undergone only minimal improvements and has not
kept pace with advancements in programming languages that could enhance
its utility. In contrast, other SQP algorithms, such as SNOPT
(\citeproc{ref-gill2005snopt}{Gill et al., 2005}), which also began
development around the same time as SLSQP, have seen continuous
improvements. SNOPT has evolved significantly through both algorithmic
enhancements and feature additions, becoming one of the leading
algorithms for nonlinear programming.

The SLSQP algorithm available in most modern packages is implemented as
a black-box function that takes the optimization functions and their
derivatives and then outputs the optimized results. These packages do
not provide users with any options for tuning the original algorithm or
for assessing the progress of an ongoing optimization. This lack of
transparency becomes a significant disadvantage for problems with
expensive optimization functions or derivatives. Users might have to
wait for hours, only to be informed at the end of the optimization
procedure that the algorithm could not converge. Several such
experiences with multiple research applications in the authors' lab were
the primary motivation behind developing the new \texttt{PySLSQP}
package.

Despite the lack of timely updates to the core algorithm and usability
improvements, SLSQP continues to be widely used in research primarily
due to its open-source nature and the availability of convenient
installation options through packages such as SciPy
(\citeproc{ref-virtanen2020scipy}{Virtanen et al., 2020}). Many
optimization practitioners use SLSQP for solving medium-sized
optimization problems with up to a hundred optimization variables and
constraints. Additionally, SLSQP is more successful compared to some of
the most advanced algorithms in solving certain classes of optimization
problems, such as optimal control problems with a coarse discretization
in time.

\begin{figure}
\centering
\includegraphics[width=0.9\textwidth,height=\textheight]{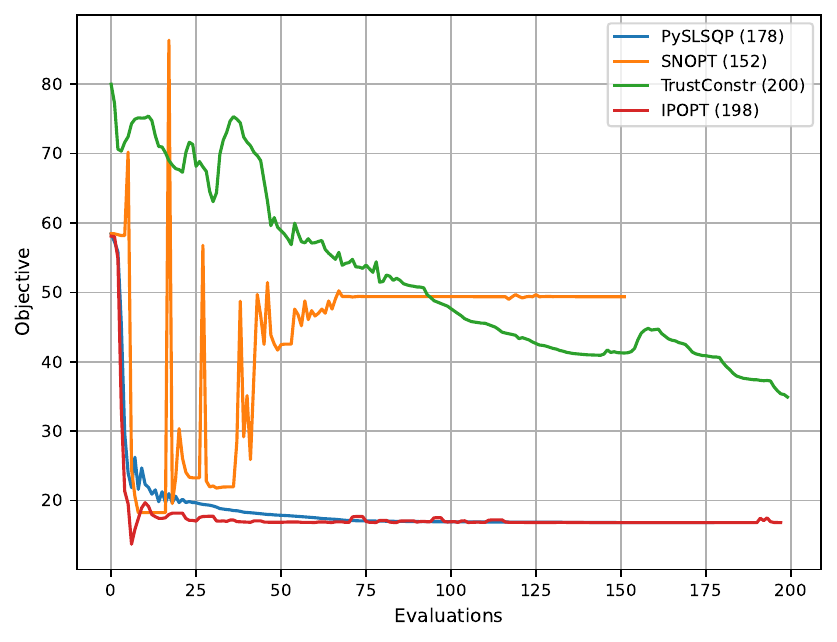}
\caption{Performance comparison for an optimal control problem.}
\end{figure}

Figure 1 above compares the convergence behaviors of \texttt{PySLSQP}
and some of the most advanced algorithms in nonlinear programming on a
coarsely discretized optimal control problem. The problem aims to
compute the optimal control parameters for a spacecraft landing
scenario. The total number of function evaluations is indicated within
parentheses in the legend. We see that \texttt{PySLSQP} is the only
algorithm that solves the problem within the 200 function evaluation
limit. Among the algorithms that failed to converge are SNOPT,
TrustConstr, and IPOPT. SNOPT is a commercial SQP algorithm, while
TrustConstr and IPOPT are Interior Point (IP) algorithms. Although IPOPT
appears to have converged in the plot, the solution returned by IPOPT
does not satisfy the feasibility criteria. This underscores the
relevance of SLSQP even today among state-of-the-art optimization
algorithms. This problem is taken from the suite of examples in the
modOpt (\citeproc{ref-modopt}{Joshy, 2024}) optimization library.

There are several optimization libraries in Python that include the
SLSQP algorithm, such as SciPy
(\citeproc{ref-virtanen2020scipy}{Virtanen et al., 2020}), NLopt
(\citeproc{ref-NLopt}{Johnson, 2024}), and pyOpt
(\citeproc{ref-perez2012pyopt}{Perez et al., 2012}). NLopt and pyOpt
require users to compile the Fortran code, which greatly deters the
majority of users from utilizing SLSQP from these libraries. pyOptSparse
(\citeproc{ref-wu2020pyoptsparse}{Wu et al., 2020}) is a fork of the
pyOpt package that supports sparse constraint Jacobians and includes
additional optimization utilities for scaling, visualization, and
storing optimization history. Most SLSQP users access it through SciPy,
which offers precompiled libraries that can be easily installed from
PyPI by running \texttt{pip\ install\ scipy}. However, like other
libraries, the SLSQP implementation in SciPy also operates as a
black-box providing limited visibility into the progress of optimization
or access to internal variables during optimization iterations. This
lack of transparency can be a drawback, particularly for users needing
more insight into the optimization process.

\texttt{PySLSQP} is developed to overcome these limitations by:

\begin{itemize}
\item
  providing a precompiled package through PyPI that can be simply
  installed with \texttt{pip\ install\ pyslsqp},
\item
  offering access to internal optimization variables at each iteration
  through a save file, and
\item
  informing users about the progress of optimization through a
  live-updated summary file and visualization.
\end{itemize}

The Python wrapper for \texttt{PySLSQP} is generated by a simple
workflow automated on GitHub, which allows even beginner developers to
tune the Fortran code for their specific application and extend the
current codebase. Offering Python-level access to internal optimization
variables such as optimality and feasibility measures, Lagrange
multipliers, etc. enables further analysis of an ongoing or completed
optimization. \texttt{PySLSQP} also features additional utilities for
numerical differentiation, scaling, warm/hot restarting, and
post-processing.

By addressing the current limitations and providing new capabilities,
\texttt{PySLSQP} enhances the transparency and usability of the SLSQP
algorithm, making it a more powerful and user-friendly tool for solving
nonlinear programming problems. \texttt{PySLSQP} is now interfaced with
the modOpt (\citeproc{ref-modopt}{Joshy, 2024}) library of optimizers,
through which it has successfully solved problems in aircraft design, as
well as aircraft and spacecraft optimal control.

\section{Software features}\label{software-features}

\subsection{Numerical differentiation}\label{numerical-differentiation}

In the absence of user-supplied first-order derivatives of the objective
or constraint functions, \texttt{PySLSQP} estimates them using
first-order finite differencing. Users have the option to set the
absolute or relative step size for the finite differences. However, it
is generally more efficient for users to provide the exact gradients, if
possible, since each finite difference estimation requires
\(\mathcal{O}(n)\) objective or constraint evaluations. Moreover, finite
difference approximations are susceptible to subtractive cancellation
errors.

\subsection{Problem scaling}\label{problem-scaling}

Scaling of the variables and functions is crucial for the convergence of
optimization algorithms. Poor scaling often leads to unsuccessful or
extremely slow optimization. \texttt{PySLSQP} enables users to scale the
optimization variables, objective, and constraints individually,
independent of the user-defined optimization functions. \texttt{PySLSQP}
automatically scales the variable bounds and derivatives according to
the user-specified scaling for the variables and functions. This allows
the user-defined initial guess, bounds, functions, and derivatives to
remain the same each time an optimization is run with a different
scaling.

\subsection{Live visualization}\label{live-visualization}

Optimization becomes slow for problems with functions or derivatives
that are costly to evaluate. In such scenarios, it is important for
users to be able to monitor the optimization process to ensure that it
is proceeding smoothly. \texttt{PySLSQP} offers the capability to
visualize the optimization progress in real-time. This feature allows
users to track convergence through optimality and feasibility measures,
and to understand how the optimization variables, objective,
constraints, Lagrange multipliers, and derivatives evolve during the
optimization. An example of a visualization generated by
\texttt{PySLSQP}, corresponding to the optimal control problem discussed
earlier, is shown in Figure 2 below.

\begin{figure}
\centering
\includegraphics{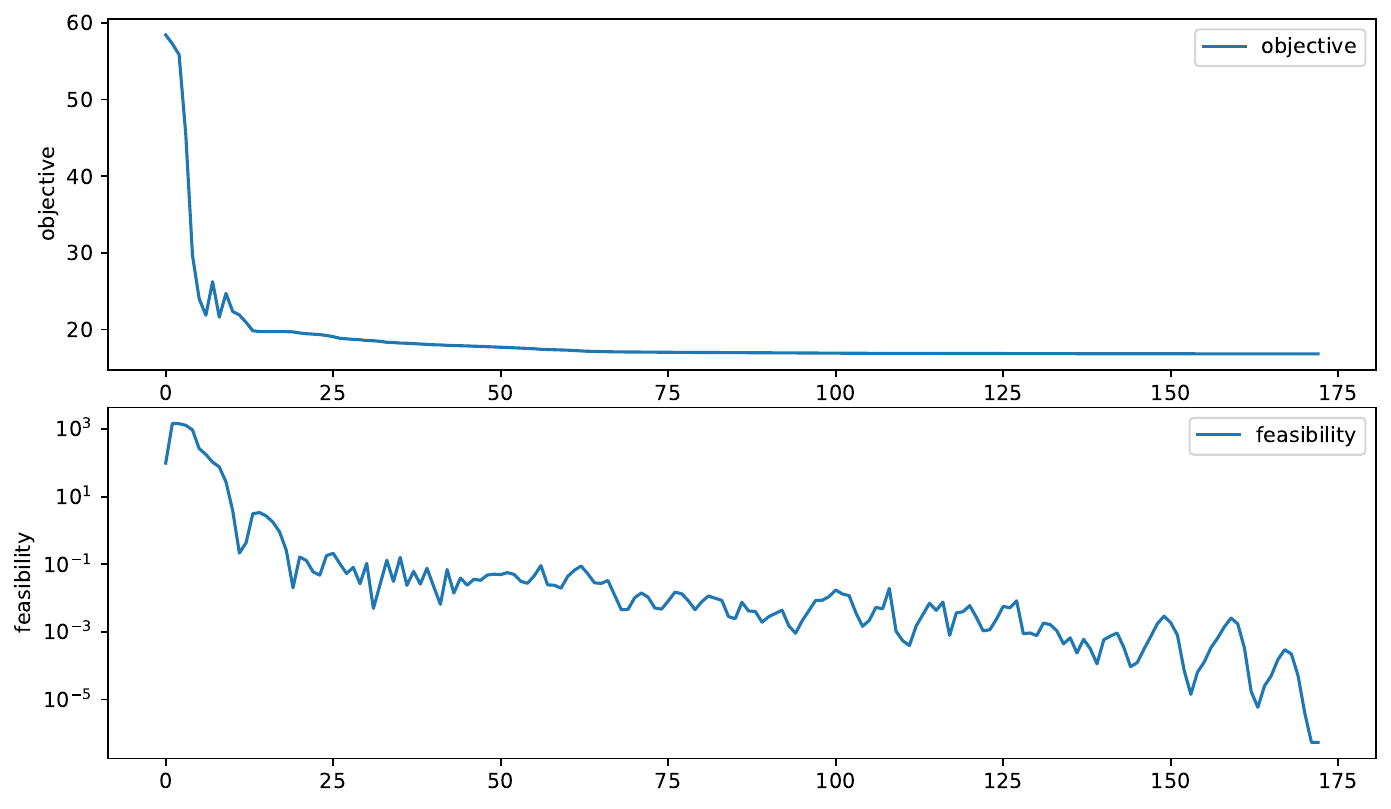}
\caption{Live visualization of the objective and feasibility.}
\end{figure}

\subsection{Access to internal optimization
data}\label{access-to-internal-optimization-data}

In addition to live visualization, \texttt{PySLSQP} provides real-time
access to optimization data through dynamically updated summary and save
files. \texttt{PySLSQP} generates a summary file that contains a table
that is updated at the end of every major iteration. This summary table
lists the values of different scalar variables in the algorithm to keep
users informed about the current state of optimization.

Users can specify which variables to save in the save file and whether
they should be saved for every iteration or only for major iterations.
The save file is valuable for analyzing optimization progress,
post-processing, or performing warm/hot restarts. It can store all
internal optimization variables - including optimization variables,
objective, constraints, objective gradient, constraint Jacobian,
optimality, feasibility, Lagrange multipliers, and line search step
sizes - facilitating advanced analysis of the optimization problem.
\texttt{PySLSQP} provides various utilities for working with data from
save files, including functions for loading and visualizing variables.

To the best of our knowledge, \texttt{PySLSQP} is the only Python
interface to the SLSQP algorithm that provides this level of access to
internal optimization information.

\subsection{Warm/Hot starting}\label{warmhot-starting}

Re-running an optimization that was terminated prematurely can be
inefficient and wasteful. For example, if a user desires higher accuracy
than was achieved in a previous run, they would need to re-execute the
optimization with a smaller accuracy parameter. Similarly, if an
optimization terminates upon reaching the iteration limit before
achieving the required accuracy, a rerun with a higher limit is
necessary to complete the process. Such repeated runs not only consume
additional computational resources but also extend the overall time
required to achieve the desired results.

To address these scenarios, \texttt{PySLSQP} offers two options for
users to efficiently restart an optimization using data from saved
files: warm starting and hot starting. In \texttt{PySLSQP}, \emph{warm
starting} refers to restarting a previously run optimization using the
most recent value of the optimization variables \(x\) from a saved file.
During a warm start, the initial guess \(x_0\) provided by the user is
replaced with the last optimization variable iterate available in the
saved file.

\emph{Hot starting} in \texttt{PySLSQP} involves re-running a previously
completed optimization by reusing the function (objective and
constraints) and derivative values from a saved file. This method is
particularly advantageous when the functions and/or their derivatives
are costly to evaluate. A significant benefit of hot starting over warm
starting is that the BFGS Hessians approximated by the SLSQP algorithm
in a hot-start will follow the same path as in the previous
optimization, while also saving the cost of function and derivative
evaluations. In contrast, during a warm start, although the algorithm
starts from the previous solution \(x^*\), the Hessian is initialized as
the identity matrix, which may necessitate more iterations to achieve
convergence.

\subsection{Ease of extension}\label{ease-of-extension}

\texttt{PySLSQP} is implemented in Python and relies on NumPy's
\emph{f2py} and the \emph{Meson} build system for compiling and
interfacing the underlying Fortran code with Python. The Python setup
script automates the build process, making it straightforward for
developers to build, install, and use \texttt{PySLSQP} after making
modifications to the Fortran code. The package also includes GitHub
workflows for automatically generating precompiled binaries in the cloud
for different system architectures using PyPA's \emph{cibuildwheel}
tool. These automated workflows ensure that \texttt{PySLSQP} remains
accessible to a broad range of users by providing consistent and
reliable installation across different platforms. Additionally, this
approach allows developers to focus on enhancing the core algorithm and
features without the overhead of managing complex build environments,
thus fostering an open-source community that can contribute effectively
to the development of the SLSQP algorithm.

\section{A simple example}\label{a-simple-example}

In this section, we solve a simple optimization problem to illustrate
some of the features explained above. In the standard SLSQP problem
format presented in the \emph{Summary}, the problem is

\begin{align*}
\underset{x \in \mathbb{R}^2}{\text{minimize}} \quad & x_1^2 + x_2^2\\
\text{subject to} \quad & x_1 + x_2 - 1 = 0, \\
& 3x_1 + 2x_2 - 1 \geq 0,  \\
\text{with} \quad & m_{eq} = 1, \; l = [0.4, -\infty]^T, \; \text{and} \; u = [+\infty, 0.6]^T.
\end{align*}

We begin by importing \texttt{numpy} and defining the optimization
functions. We will only define the derivatives for the constraints and
let \texttt{PySLSQP} approximate the derivatives for the objective
function. We then define the constants for the optimization, which
include the variable bounds, number of equality constraints, initial
guess, and scaling factors.

\begin{Shaded}
\begin{Highlighting}[]
\ImportTok{import}\NormalTok{ numpy }\ImportTok{as}\NormalTok{ np}

\KeywordTok{def}\NormalTok{ objective(x):}
    \ControlFlowTok{return}\NormalTok{ x[}\DecValTok{0}\NormalTok{]}\OperatorTok{**}\DecValTok{2} \OperatorTok{+}\NormalTok{ x[}\DecValTok{1}\NormalTok{]}\OperatorTok{**}\DecValTok{2}

\KeywordTok{def}\NormalTok{ constraints(x):}
    \ControlFlowTok{return}\NormalTok{  np.array([x[}\DecValTok{0}\NormalTok{] }\OperatorTok{+}\NormalTok{ x[}\DecValTok{1}\NormalTok{] }\OperatorTok{{-}} \DecValTok{1}\NormalTok{, }\DecValTok{3}\OperatorTok{*}\NormalTok{x[}\DecValTok{0}\NormalTok{] }\OperatorTok{+} \DecValTok{2}\OperatorTok{*}\NormalTok{x[}\DecValTok{1}\NormalTok{] }\OperatorTok{{-}} \DecValTok{1}\NormalTok{])}

\KeywordTok{def}\NormalTok{ jacobian(x):}
    \ControlFlowTok{return}\NormalTok{ np.array([[}\DecValTok{1}\NormalTok{, }\DecValTok{1}\NormalTok{], [}\DecValTok{3}\NormalTok{, }\DecValTok{2}\NormalTok{]])}

\CommentTok{\# Variable bounds}
\NormalTok{x\_lower }\OperatorTok{=}\NormalTok{ np.array([}\FloatTok{0.4}\NormalTok{, }\OperatorTok{{-}}\NormalTok{np.inf])}
\NormalTok{x\_upper }\OperatorTok{=}\NormalTok{ np.array([np.inf, }\FloatTok{0.6}\NormalTok{])}

\CommentTok{\# Number of equality constraints}
\NormalTok{m\_eq }\OperatorTok{=} \DecValTok{1}

\CommentTok{\# Initial guess}
\NormalTok{x0 }\OperatorTok{=}\NormalTok{ np.array([}\DecValTok{2}\NormalTok{,}\DecValTok{3}\NormalTok{])}

\CommentTok{\# Scaling factors}
\NormalTok{x\_s }\OperatorTok{=} \FloatTok{10.0}
\NormalTok{o\_s }\OperatorTok{=}  \FloatTok{2.0}
\NormalTok{c\_s }\OperatorTok{=}\NormalTok{ np.array([}\FloatTok{1.}\NormalTok{, }\FloatTok{0.5}\NormalTok{])}
\end{Highlighting}
\end{Shaded}

Most of the features in \texttt{PySLSQP} are accessed through the
\texttt{optimize} function. We now import \texttt{optimize} and solve
the problem by calling it with the functions and constants defined
above. When calling \texttt{optimize}, we will define the absolute step
size for the finite difference approximation of the objective gradient.
Additionally, we instruct \texttt{PySLSQP} to save the optimization
variables \(x\) and the objective value \(f\) from each major iteration
to a file named \texttt{save\_file.hdf5}. Lastly, we configure the
arguments to live-visualize the objective \(f\) and the variable \(x_1\)
during the optimization.

\begin{Shaded}
\begin{Highlighting}[]
\ImportTok{from}\NormalTok{ pyslsqp }\ImportTok{import}\NormalTok{ optimize}

\NormalTok{results }\OperatorTok{=}\NormalTok{ optimize(x0, obj}\OperatorTok{=}\NormalTok{objective, con}\OperatorTok{=}\NormalTok{constraints, jac}\OperatorTok{=}\NormalTok{jacobian, }
\NormalTok{                   meq}\OperatorTok{=}\NormalTok{m\_eq, xl}\OperatorTok{=}\NormalTok{x\_lower, xu}\OperatorTok{=}\NormalTok{x\_upper, finite\_diff\_abs\_step}\OperatorTok{=}\FloatTok{1e{-}6}\NormalTok{,}
\NormalTok{                   x\_scaler}\OperatorTok{=}\NormalTok{x\_s, obj\_scaler}\OperatorTok{=}\NormalTok{o\_s, con\_scaler}\OperatorTok{=}\NormalTok{c\_s,}
\NormalTok{                   save\_itr}\OperatorTok{=}\StringTok{\textquotesingle{}major\textquotesingle{}}\NormalTok{, save\_vars}\OperatorTok{=}\NormalTok{[}\StringTok{\textquotesingle{}majiter\textquotesingle{}}\NormalTok{, }\StringTok{\textquotesingle{}x\textquotesingle{}}\NormalTok{, }\StringTok{\textquotesingle{}objective\textquotesingle{}}\NormalTok{],}
\NormalTok{                   save\_filename}\OperatorTok{=}\StringTok{"save\_file.hdf5"}\NormalTok{,}
\NormalTok{                   visualize}\OperatorTok{=}\VariableTok{True}\NormalTok{, visualize\_vars}\OperatorTok{=}\NormalTok{[}\StringTok{\textquotesingle{}objective\textquotesingle{}}\NormalTok{, }\StringTok{\textquotesingle{}x[0]\textquotesingle{}}\NormalTok{])}

\CommentTok{\# Print the returned results dictionary}
\BuiltInTok{print}\NormalTok{(results)}
\end{Highlighting}
\end{Shaded}

Once \texttt{optimize} is executed, a summary of the optimization will
be printed to the console. The function also returns a dictionary that
contains the results of the optimization. By default, \texttt{PySLSQP}
writes the summary of the major iterations to a file named
\texttt{slsqp\_summary.out}.

For additional usage guidelines, API reference, and installation
instructions, please consult the
\href{https://pyslsqp.readthedocs.io/}{documentation}.

\section{Acknowledgements}\label{acknowledgements}

This work was supported by NASA under award No.~80NSSC23M0217.

\section*{References}\label{references}
\addcontentsline{toc}{section}{References}

\phantomsection\label{refs}

\end{document}